\UseRawInputEncoding
\documentclass[%
 aip,
 amsmath,amssymb,
reprint,%
]{revtex4-1}

\usepackage{graphicx}
\usepackage{dcolumn}
\usepackage{bm}

\usepackage[utf8]{inputenc}
\usepackage[T1]{fontenc}
\usepackage{mathptmx}
\usepackage{etoolbox}
\usepackage[ngerman, english]{babel}
\usepackage{xcolor}

\newcommand{\degC}{\textdegree C\;}
\newcommand{\VO}{V$_O^{\bullet \bullet}$}

\DeclareUnicodeCharacter{2212}{-}

\makeatletter
\def\@email#1#2{%
 \endgroup
 \patchcmd{\titleblock@produce}
  {\frontmatter@RRAPformat}
  {\frontmatter@RRAPformat{\produce@RRAP{*#1\href{mailto:#2}{#2}}}\frontmatter@RRAPformat}
  {}{}
}%
\makeatother
\begin{document}

\preprint{AIP/123-QED}

\title[]{Influence of charged walls and defects on DC resistivity and dielectric relaxations in Cu-Cl boracite}
\author{C. Cochard}
 \affiliation{School of Science and Engineering, University of Dundee, Nethergate, Dundee, DD1 4HN, United Kingdom}
 \altaffiliation[Also at ]{School of Mathematics and Physics, Queen’s University Belfast, Belfast, BT7 1NN, United Kingdom.}
 \email{CCochard001@dundee.ac.uk}
\author{T. Granzow}%
 \affiliation{ 
MRT Department, Luxembourg Institute of Science and Technology (LIST), L-4362 Esch-sur-Alzette, Luxembourg
}%

\author{C. M. Fernandez-Posada}
 \affiliation{Department of Earth Sciences, University of Cambridge, Downing Street, Cambridge CB2 3EQ, United Kingdom}%
 \altaffiliation[Also at ]{Maxwell Centre, Cavendish Laboratory, University of Cambridge, Cambridge, United Kingdom}
 
\author{M. A. Carpenter}
 \affiliation{Department of Earth Sciences, University of Cambridge, Downing Street, Cambridge CB2 3EQ, United Kingdom}%
\author{R. G. P. McQuaid}
 \affiliation{School of Mathematics and Physics, Queen’s University Belfast, Belfast, BT7 1NN, United Kingdom}%
\author{J. M. Guy}
 \affiliation{School of Mathematics and Physics, Queen’s University Belfast, Belfast, BT7 1NN, United Kingdom}%
\author{R. W. Whatmore}
 \affiliation{Department of Materials, Faculty of Engineering, Imperial College London, London, SW7 2AZ, United Kingdom}%
 
\author{J. M. Gregg}
 \affiliation{School of Mathematics and Physics, Queen’s University Belfast, Belfast, BT7 1NN, United Kingdom}%

\date{\today}

\begin{abstract}
Charged domain walls form spontaneously in Cu-Cl boracite on cooling through the phase transition. These walls exhibit changed conductivity compared to the bulk and motion consistent with the existence of negative capacitance. Here, we present the dielectric permittivity and DC resistivity of bulk Cu-Cl boracite as a function of temperature (-140\,\textdegree C to 150\,\textdegree C) and frequency (1\,mHz to 10\,MHz). The thermal behaviour of the two observed dielectric relaxations and the DC resistivity is discussed.  We propose that the relaxations can be explained by the existence of point defects, most likely local complexes created by a change of valence of Cu and accompanying oxygen vacancies. In addition, the sudden change in resistivity seen at the phase transition suggests that conductive domain walls contribute significantly to the conductivity in the ferroelectric phase.
\end{abstract}

\begin{widetext}
This article may be downloaded for personal use only. Any other use requires prior permission of the author and AIP Publishing. This article appeared in Appl. Phys. Lett. \textbf{119} 202904 (2021)  and may be found at (\url{http://doi.org/10.1063/5.0067846})
\end{widetext}

\maketitle

\begin{figure}
\includegraphics[width=0.9\linewidth]{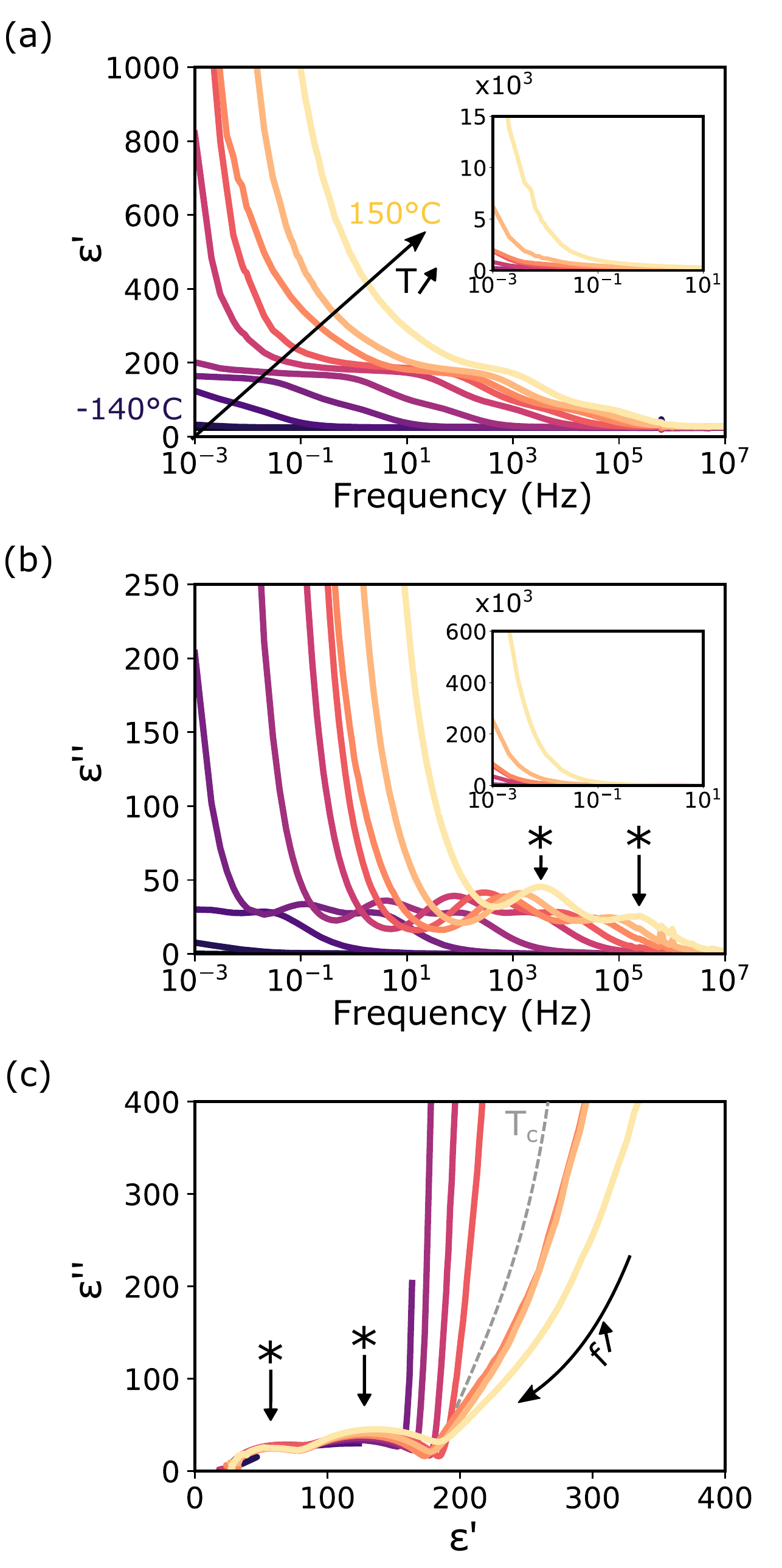}
\caption{\label{fig:diel} Frequency dependence of the dielectric permittivity
(a) real part, (b) imaginary part and (c) Cole-Cole plot. * highlight the two relaxations in (b) and (c)
The colour lines are separated by 20\,\degC except close to the phase transition where they are separated by 10\,\degC.}
\end{figure}

\begin{figure}
\includegraphics[width=\linewidth]{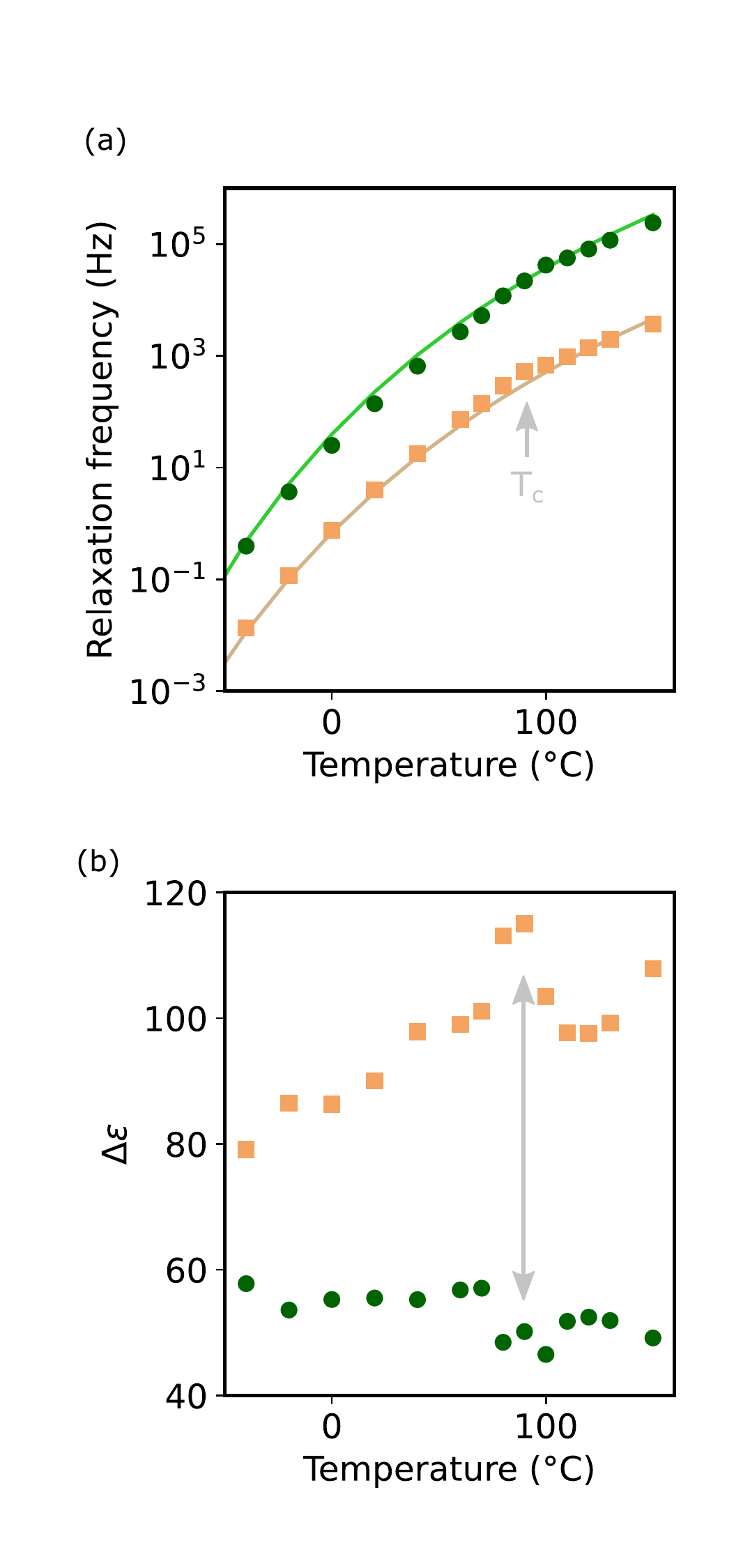}
\caption{\label{fig:fit} Fitting parameters of relaxations 
(a) relaxation frequency, (b) dielectric strength. The dark green circles represent the high frequency relaxation and the light orange squares the lower frequency one.}
\end{figure}

\begin{figure}
\includegraphics[width=\linewidth]{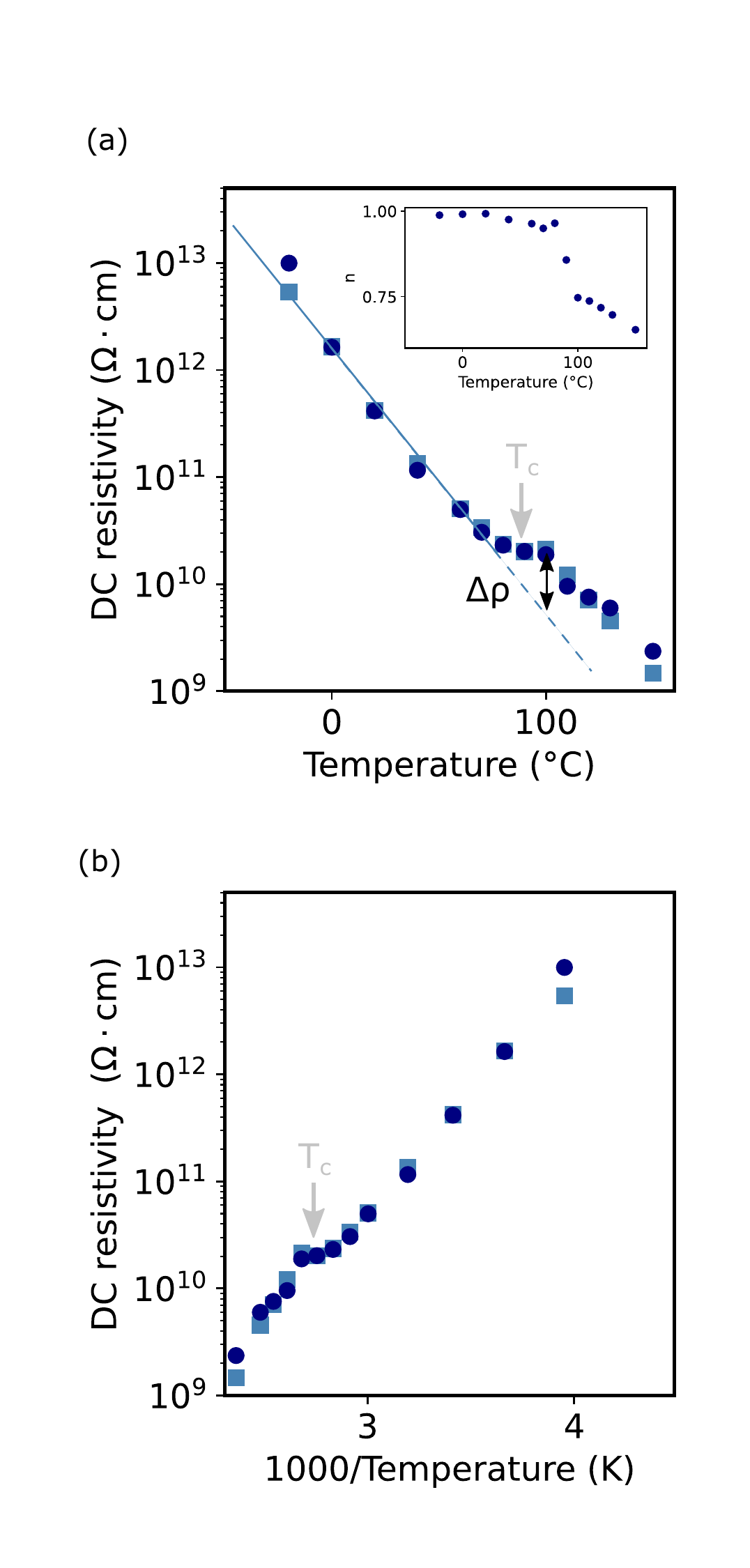}
\caption{\label{fig:res} Thermal evolution of resistivity
(a) as a function of temperature (b) as a function of the inverse of temperature. Light blue squares represent the value of resistivity measured at 1mHz. Dark blue circles depict the value obtained from fitting. The exponent n is presented in the inset of (a). 
}
\end{figure}

The boracites form a class of ferroelectrics with the general formula M$_3$B$_7$O$_{13}$X, where M is a metal and X a halide.  As improper ferroelectrics, their interesting crystallography and phase transitions attracted attention in the 1960’s, 70's and 80's with the first coupled magnetoelectric multiferroic switching being demonstrated by Ascher et al.\cite{Ascher_Rieder_Schmid_Stossel_1966} in Ni$_3$B$_7$O$_{13}$I and with observations indicative of potential for electro-optic\cite{Schmid_SchwarzmuLler_1976} and pyroelectric applications\cite{Schmid_Genequand_Pouilly_Chan_1980, Smith_Rosar_Shaulov_1981, Whatmore_Herbert_Ainger_1980}.  However, the growth of large single crystals\cite{Schmid_1965, Whatmore_Brierley_Ainger_1980} is difficult and with the development of perovskite oxide ceramic materials possessing diverse functional properties, interest in the boracite family subsequently waned. Recently, however, the discoveries surrounding improper ferroelectrics\cite{Catalan_Seidel_Ramesh_Scott_2012, Evans_Cochard_McQuaid_Cano_Gregg_Meier_2020, Feng_Xu_Bellaiche_Xiang_2018} and their associated charged domain walls \cite{Guy2021, McQuaid_Campbell_Whatmore_Kumar_Gregg_2017} has rekindled interest in the potential of boracites as functional materials. The charged domain walls in Cu-Cl boracite are particularly remarkable: they present either enhanced conductivity (in 90\textdegree\; tail-to-tail walls) or reduced conductivity (in 90\textdegree\; head-to-head walls) relative to the bulk\cite{McQuaid_Campbell_Whatmore_Kumar_Gregg_2017}. Charged walls exist spontaneously in Cu-Cl boracite, to accommodate for the spontaneous shear strain developing at the phase transition, but they can also be injected and repositioned by stress and electric field, making them interesting for the future of nanoelectronics\cite{Seidel_2012, Whyte_Gregg_2015}. Uniquely, head-to-head charged walls have been shown to have an unconventional electrostatic response (moving in the opposite direction to that expected under an applied electric field), consistent with the existence of negative capacitance\cite{Guy2021}. This discovery of new properties at domain walls prompts our understanding of the intrinsic properties of boracites to be revisited. \\
The characterisation of the dielectric properties is particularly important, as it pertains to the unusual electrostatic response of charged domain walls\cite{Guy2021}, as well as playing a role in the piezoelectric and ferroelectric responses. Here, we focus on the characterisation of the dielectric dispersion and resistivity, as functions of temperature, of the same Cu-Cl boracite sample in which the conductivity of charged walls\cite{McQuaid_Campbell_Whatmore_Kumar_Gregg_2017} and negative capacitance were measured\cite{Guy2021}. We observe two dielectric relaxations and a strong change of resistivity at the phase transition (90\,\textdegree C) between the high-temperature piezoelectric phase and the low-temperature ferroelectric phase\cite{Fernandez-Posada_Cochard_Gregg_Whatmore_Carpenter_2021}. \\
The Cu$_3$B$_7$O$_{13}$Cl single-crystal was prepared by the phase transport technique\cite{Whatmore_Brierley_Ainger_1980}. The crystal is about 5x5x1mm$^3$ mirror polished with faces parallel. The sample is transparent with a faint blue coloration. The transmission spectrum of a small piece of the sample was measured on a PerkinElmer Lambda900  spectrometer between 300\,nm and 900\,nm. The impedance of this crystal was measured at frequencies between 1\,mHz and 10\,MHz and a driving voltage of 0.5V in a large temperature range of -140\,\degC to 150\,\degC, spanning the piezoelectric-ferroelectric phase transition, with steps of 10 or 20\,\degC using a Novocontrol Concept 40 dielectric spectrometer with the unpoled sample directly contacted with parallel brass plates freshly polished to ensure good contacts and mounted with a mechanical spring. The complex dielectric permittivity $\varepsilon^*=\varepsilon'+i\varepsilon''$ and resistivity $\rho^*=\rho'+i\rho''$ were calculated classically from the complex impedance. \\
Figure \ref{fig:diel} presents the frequency dispersion of $\varepsilon'$ and $\varepsilon''$ at different temperatures. Piezoelectric resonances are observed at frequencies above 500\,kHz. With decreasing frequency, two relaxations can be clearly observed at frequencies around 1\,kHz and 100\,kHz at 150\,\textdegree C. Upon further lowering of frequencies, the real and imaginary parts of the dielelectric permittivity increase rapidly. This increase is consistent with the existence of a resistor in series.  
The real and imaginary part were fitted simultaneously for frequencies below piezoelectric resonances for all temperatures. The relaxations were modelled as a Cole-Cole relaxation, and a term was added to describe the low-frequency conductive behaviour, as is commonly done in ferroelectric materials\cite{Granzow_2017}
\begin{eqnarray}
    \label{eq:fit}
    \varepsilon ^*=\frac{\Delta \varepsilon_1}{1+i(f/f_{r,1})^{a_1}}+\frac{\Delta \varepsilon_2}{1+i(f/f_{r,2})^{a_2}}+\frac{\sigma}{\varepsilon_0(i2\pi f)^n}+\varepsilon _\infty
\end{eqnarray}
where $\Delta \varepsilon_1$ and $\Delta \varepsilon_2$ are the amplitudes of the relaxations, $f_{r,1}$ and $f_{r,2}$ the two characteristic relaxation frequencies, $a_1$ and $a_2$ two exponents, $\sigma$ is the DC conductivity, $n$ is an empirical constant and $\varepsilon_\infty$ is the permittivity at high frequencies. Fitting the two relaxations with the empirical Havriliak-Negami function\cite{Havriliak_Negami_1967} improves marginally the quality of the fit, whereas fitting with two simple Debye relaxations decreases the quality of the fit with respect to the conduction-dominated low-frequency part. All fits lead to similar results regarding the changes in amplitude and frequencies of the relaxations (Supplementary Fig. S1 and S2). The results of fitting at low temperatures (below -50\,\degC) was not considered since the two relaxations have moved towards low frequencies and it becomes difficult to differentiate them from one another and from the low frequency conduction. The fitting parameters for the two relaxations, on the one hand, and the DC conductivity, on the other hand, are presented in Fig.\:\ref{fig:fit} and Fig.\;\ref{fig:res}, respectively, and will be discussed separately.\\
Figure\;\ref{fig:fit} presents the fitted parameters using the Cole-Cole model for the two relaxations. Both relaxations have similar thermal evolution: the relaxation frequencies (Fig.\;\ref{fig:fit}a) continuously increase with increasing temperature, the amplitudes of the two relaxations display a jump at the phase transition(Fig.\;\ref{fig:fit}b)  and the exponents $a_1$ and $a_2$ remain equal to 0.8 across the temperature range investigated (Supplementary Fig. S2). \\
The smooth continuous increase in relaxation frequencies is a strong indication that these relaxations are not related to Maxwell-Wagner polarisation. Indeed, the frequency associated with this type of relaxation are inversely proportional to the resistivity and dielectric permittivity, and both of these quantities display a sharp increase at the phase transition. This would lead to a sharp decrease of the relaxation frequencies at the phase transition which is absent here.\\
The temperature evolution of relaxation frequencies is often modelled using the Vogel-Fulcher law, describing an activated process of activation energy $E_a$ and freezing below the temperature $T_f$
\begin{eqnarray}
    f=f_0e^{-\frac{E_a}{k_B(T-T_f)}}
\end{eqnarray}
where $k_B$ is the Boltzmann constant and $f_0$ the attempt frequency. The continuous lines in Fig. \ref{fig:fit}(a) represent the results of the fit to the measured data. Fitting a Vogel-Fulcher law is never trivial\cite{Cochard_Bril_Guedes_Janolin_2016}, but a few things can be said based on the fitted values shown in Table~\ref{tab:VF}. \\
The freezing temperatures were found to be $\sim$\,0\,K (within error), suggesting that the processes leading to the relaxations will not freeze. Fixing the freezing temperature to 0\,K, i.e. assuming an Arrhenius law, leads to similar results with activation energies that are comparable to the one observed for the lower frequency relaxation, but still within the error associated with the Vogel-Fulcher fits. \\
A recent study\cite{Fernandez-Posada_Cochard_Gregg_Whatmore_Carpenter_2021}, focusing on the elastic properties of this sample, identified a relaxation process freezing at $\sim40$\,K, which was attributed to strain relaxation around local ferroelectric dipoles or polarons. The low freezing temperature observed could be broadly consistent with our observations; however, the minimum activation energy ($\sim\,0.02$\,eV) associated with the strain process is more than an order of magnitude lower than that observed in this work. The differences in activation energies and freezing behaviour suggest different microscopic origins for the relaxation processes reported in the present work and the one previously observed. Specifically, the absence of freezing of the relaxation observed here suggests that the reorientations of local dipoles are independent. Indeed, in a process modelled by an Arrhenius law, there is no assumption of collective response to the applied electric field.

\begin{table}
\caption{Fitting parameters of the Vogel-Fulcher analysis. Relaxation 1 and 2 correspond to the relaxation occurring at the highest and lowest frequency, respectively.}
\begin{ruledtabular}
\begin{tabular}{cccc}
\label{tab:VF}
&$T_f (K) $&$E_a$\,(eV)&$f_0$\,(Hz)\\
\hline
Relaxation 1 & $-9\pm13$ & $0.59\pm0.06$ & $(8.0\pm10.0)\cdot 10^{10}$\\
Relaxation 2 & $-1\pm10$ & $0.64\pm0.05$ & $(1.8\pm5.3)\cdot 10^{13}$
\end{tabular}
\end{ruledtabular}
\end{table}

\begin{figure}
\includegraphics[width=\linewidth]{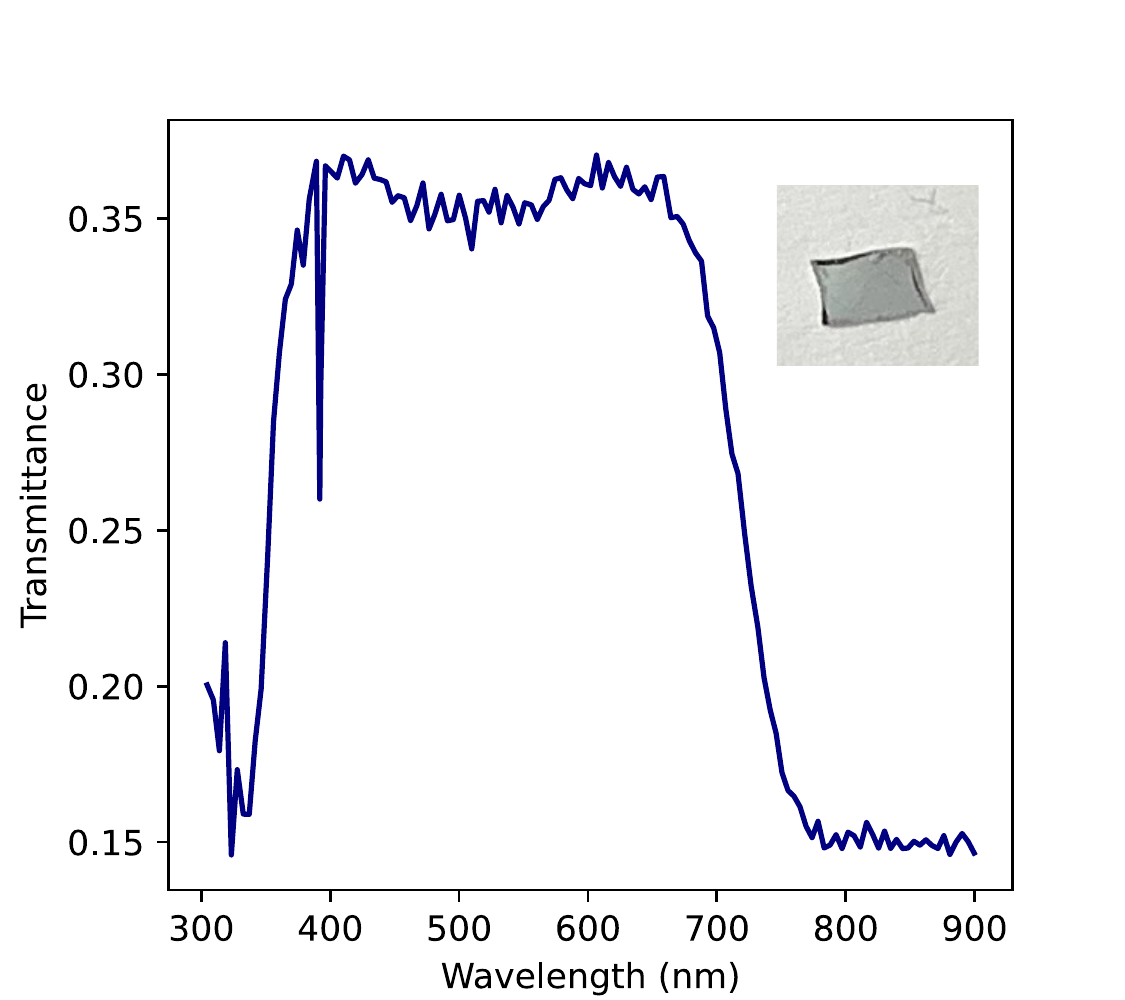}
\caption{\label{fig:uv-vis} Transmittance of the Cu-Cl boracite sample (photo as inset)   
}
\end{figure}

Notably, the activations energies for both relaxations are the same within error ($\sim\,0.6$\,eV). This suggests a common origin. Domain wall motion can largely be excluded, as both relaxation phenomena persist above $T_c$. While a contribution from domains/domain walls below $T_c$ cannot be completely excluded, it is certainly not the main relaxation phenomenon at play.\\
Perhaps the most classical origin of dielectric relaxations in dielectrics, point defects\cite{Elissalde_Ravez_2001, Nataf_Aktas_Granzow_Salje_2016, Pramanick_Prewitt_Forrester_Jones_2012} are the most likely mechanism for the observed dielectric relaxations. Indeed, the order of magnitude of the activation energies is very similar to that reported for defect states in perovskites\cite{Akkopru-Akgun_2019}. UV-visible optical spectroscopy (Fig.\;\ref{fig:uv-vis}) indicates the existence of states in the band gap, strongly supporting point defects as a source of the dielectric relaxations. The colour of the sample and the band in the transmission spectrum at 490-540nm\cite{Kim_Somoano_LoweMa_Coleman_Moopen_1981} both suggest a change of valence of copper Cu$^{2+}\rightarrow \textrm{Cu}^+$. To conserve charge neutrality, this change of valence is likely to be associated with positively charged point defects, such as oxygen vacancies V$_O^{\bullet \bullet}$.\\
Point defects can lead to dielectric relaxations in several ways: creating a local dipole (or bound charge) that can be reoriented by the electric field, providing a free charge carrier that is moved by the electric field or through the coupling to phonons \cite{Gridnev2002} In parenthesis, defect mobility leads to dielectric relaxations in some cases, but have usually much higher activation energies ($\sim$\,2\,eV in boracite glass\cite{El-Falaky_Guirguis_AbdEl-Aal_2012}) and are rarely observed in ferroelectrics around room temperature. The similar activation energies suggest that the two relaxations “experience” a similar energy landscape. On the other hand, the difference in $f_0$ indicates that, at infinite temperature, the energy barrier between the minima is overcome at different frequencies. This could be explained with different effective masses of the dipoles or charges or by the coupling to two different phonon modes.\\
For example, the change of valence in copper creates a local dipole but can also lead to polaronic hopping between sites. These two mechanisms would probably have similar energy landscapes, but the dipole reorientation and polaron hopping would have different effective masses. Alternatively, the existence of anisotropy in some of the B-FO tetrahedra could create relaxations with different effective masses. Indeed, in half of the tetrahedra, boron is not located at the centre. This off-centring leads to the existence of local dipoles (even in the high temperature phase) pointing towards chlorine. The position of a \VO vacancy would have a strong influence on this local B-O dipole and in turn change the effective mass of the dipole created directly by \VO. A third option would be a defect complex formed of Cu$^+$  and \VO; this seems particularly likely, in terms of keeping the electroneutrality of the sample. This complex would have one activation energy, but the \mbox{Cu$^+$- dipoles} and \VO would again have different effective masses. In this discussion, the polarisability of the defects was the main mechanism considered; however, defects can also lead to dielectric relaxation through the coupling to phonon modes. We believe that this coupling is less likely in the present case as the phase transition does not affect the temperature dependence of the relaxation frequency, whereas phonons would be strongly affected. At this point and 
by analogy with acceptor- or donor-doped perovskite\cite{Sapper_Dittmer_Damjanovic_Erdem_Keeble_Jo_Granzow_Rodel_2014, Zhang_Erdem_Ren_Eichel_2008} and the observed decrease in formation energy when a defect complex is created\cite{Eichel_Erhart_Traskelin_Albe_Kungl_Hoffmann_2008, Erhart_Eichel_Traskelin_Albe_2007}, we consider that a charge transfer between the two defects is the most likely phenomenon. \\
Below the frequencies of the dielectric relaxations, a strong increase in both $\varepsilon'$ and $\varepsilon''$ is observed. It is attributed to the DC conductivity of the sample, which can be determined in two different ways: (i) as a result of fitting to the dielectric dispersion data and (ii) considering the low-frequency (1\,mHz) value of the resistivity $\rho$ (Supplementary Figure 
S3a). In both cases, the temperature range was limited to -50\,\degC and above. Figure \ref{fig:res} presents the thermal behaviour of resistivity determined both ways. The values are in excellent agreement and are of the order of magnitude of the resistivity previously reported in Cu-Cl boracite\cite{Schmid_Petermann_1977}, which suggests good reliability of our data.\\
At the phase transition, a clear jump in resistivity ($\Delta\rho$) of about an order of magnitude is observed. A similar step was seen in the exponent n, determined in the fitting procedure (see inset of Fig. 3(a)). This abrupt change is not an artefact, since it can be observed directly on the Cole-Cole plot (Fig. \ref{fig:diel}c): there is a clear and sudden change in the low frequency impedance across the phase transition. Schmid and Petermann\cite{Schmid_Petermann_1977} also reported a sudden change in resistivity at the phase transition, although they observed that the sign of $\Delta\rho$ depended on the sample and the poling state. This led the authors, in 1977, to conclude that “this may possibly be due to some spurious domain orientation […] or to a higher conduction along the walls”, even though domain wall conduction had not been experimentally observed in any ferroelectric system at that time. We observed a similar influence of the domain pattern in the DC resistivity measured at room temperature for the same sample after different thermal treatment (Supplementary Fig. S2b and S4): the highest resistivity being more than twice as large as the smallest one. However, now  that enhanced or depressed conduction at domain walls\cite{Guy2021, McQuaid_Campbell_Whatmore_Kumar_Gregg_2017} has been previously observed experimentally through spatially resolved current mapping, it is more likely that the domain walls influence predominantly the DC resistivity rather than “spurious domains”, since single domains samples consistently show higher resistivity than multidomain ones\cite{Schmid_Petermann_1977}.\\
In addition to its abrupt change at the phase transition, the DC resistivity decreases with increasing temperature, which indicates an insulator/semiconductor behaviour. It is common to study the temperature dependence to get insight into the physical mechanism responsible for carrier transport.  Since Cu-Cl boracite has an optical band gap of 4\,eV and the observation of defects state in the band gap, a change in resistivity due to activated hopping between potential wells seems the most likely mechanism. In this case, resisitivity, follows an Arrhenius law:
 \begin{eqnarray}
     \rho_{DC}=\rho_0e^\frac{E_a}{k_BT}
 \end{eqnarray}
where $\rho_0$ is a prefactor representing either the resistivity at at infinite temperature, $E_a$ the activation energy and $k_B$ the Boltzmann constant.\\
The activation energies were found to differ on each side of the phase transition with $E_a^{\rho}=0.42\pm0.04$\,eV below and $E_a^{\rho}=0.72\pm0.08$\,eV above the transition, respectively. These values of activation energies are of the same order of magnitude as the values reported in the literature for Cu-Cl boracite, although we observe the higher activation energy above $T_C$. Additionally, they are also of the same order as the one reported for the temperature dependence of the electronic conductivity driven by charge transfer between Cu$^+$  and Cu$^{2+}$ in a halide double perovskite\cite{Connor_Smaha_Li_Gold-Parker_Heyer_Toney_Lee_Karunadasa_2021}.\\
The thermal dependences of the resisitivity are of the same order of magnitude as the activation energies compared to the the one observed for the two dielectric relaxations: it would be natural to assume a similar origin. To explain the difference on each side of the phase transition, it could be hypothesized (see Supplementary Note 1) that, below the phase transition, the decrease in activation energy is due to the existence of conducting domain walls, that tend to display intrinsic semiconducting behaviour\cite{Eliseev_Morozovska_Svechnikov_Gopalan_Shur_2011, Liu_Zheng_Koocher_Takenaka_Wang_Rappe_2015, Nataf_Guennou_Gregg_Meier_Hlinka_Salje_Kreisel_2020}.\\
In summary, the impedance spectra of Cu-Cl boracite were measured at different temperatures. Two dielectric relaxations and an increase of permittivity at low frequencies, consistent with the existence of a series resistor, were found.  The study of the temperature evolution of the two relaxations and the low-frequency conductivity revealed the role of point defects. Based on the similarity of activation energies ($\sim$\,0.6\,eV) and differences in attempt frequencies ($f_0=7-9\cdot 10^{12}$\,Hz), we suggest that defect complexes, related to a change of valence of Cu and oxygen vacancies, are the most likely. Below the ferroelectric phase transition temperature, a decrease in resistivity is observed, consistent with the appearance of the conductive domain walls known in this system.  

\section*{Supplementary Material}
See supplementary material for the difference between Debye and Havriliak-Negami fitting, the effect of thermal treatment on the complex dielectric permittivity and the resistivity dispersion.

\begin{acknowledgments}
The assistance of C. J. Brierley at Plessey Research (Caswell) Ltd. in growing the boracite crystals is gratefully acknowledged. The authors acknowledge funding from the Engineering and Physical Sciences Research Council (EPSRC: EP/P02453X/1; EP/P020194/1), the US-Ireland Research and Development Partnership Programme (USI 120) 
\end{acknowledgments}

\section*{ Author Declarations}
\subsection*{Conflict of interest}
The authors have no conflicts to disclose.

\section*{Data Availability Statement}
The data that support the findings of this study are available from the corresponding author upon reasonable request.

\nocite{*}
\bibliography{boracite_diel}

\begin{thebibliography}{34}%
\makeatletter
\providecommand \@ifxundefined [1]{%
 \@ifx{#1\undefined}
}%
\providecommand \@ifnum [1]{%
 \ifnum #1\expandafter \@firstoftwo
 \else \expandafter \@secondoftwo
 \fi
}%
\providecommand \@ifx [1]{%
 \ifx #1\expandafter \@firstoftwo
 \else \expandafter \@secondoftwo
 \fi
}%
\providecommand \natexlab [1]{#1}%
\providecommand \enquote  [1]{``#1''}%
\providecommand \bibnamefont  [1]{#1}%
\providecommand \bibfnamefont [1]{#1}%
\providecommand \citenamefont [1]{#1}%
\providecommand \href@noop [0]{\@secondoftwo}%
\providecommand \href [0]{\begingroup \@sanitize@url \@href}%
\providecommand \@href[1]{\@@startlink{#1}\@@href}%
\providecommand \@@href[1]{\endgroup#1\@@endlink}%
\providecommand \@sanitize@url [0]{\catcode `\\12\catcode `\$12\catcode
  `\&12\catcode `\#12\catcode `\^12\catcode `\_12\catcode `\%12\relax}%
\providecommand \@@startlink[1]{}%
\providecommand \@@endlink[0]{}%
\providecommand \url  [0]{\begingroup\@sanitize@url \@url }%
\providecommand \@url [1]{\endgroup\@href {#1}{\urlprefix }}%
\providecommand \urlprefix  [0]{URL }%
\providecommand \Eprint [0]{\href }%
\providecommand \doibase [0]{http://dx.doi.org/}%
\providecommand \selectlanguage [0]{\@gobble}%
\providecommand \bibinfo  [0]{\@secondoftwo}%
\providecommand \bibfield  [0]{\@secondoftwo}%
\providecommand \translation [1]{[#1]}%
\providecommand \BibitemOpen [0]{}%
\providecommand \bibitemStop [0]{}%
\providecommand \bibitemNoStop [0]{.\EOS\space}%
\providecommand \EOS [0]{\spacefactor3000\relax}%
\providecommand \BibitemShut  [1]{\csname bibitem#1\endcsname}%
\let\auto@bib@innerbib\@empty
\bibitem [{\citenamefont {Ascher}\ \emph {et~al.}(1966)\citenamefont {Ascher},
  \citenamefont {Rieder}, \citenamefont {Schmid},\ and\ \citenamefont
  {Stössel}}]{Ascher_Rieder_Schmid_Stossel_1966}%
  \BibitemOpen
  \bibfield  {author} {\bibinfo {author} {\bibfnamefont {E.}~\bibnamefont
  {Ascher}}, \bibinfo {author} {\bibfnamefont {H.}~\bibnamefont {Rieder}},
  \bibinfo {author} {\bibfnamefont {H.}~\bibnamefont {Schmid}}, \ and\ \bibinfo
  {author} {\bibfnamefont {H.}~\bibnamefont {Stössel}},\ }\bibfield  {title}
  {\enquote {\bibinfo {title} {{Some Properties of Ferromagnetoelectric
  Nickel‐Iodine Boracite,
  $\textrm{Ni}_3\textrm{B}_7\textrm{O}_{13}\textrm{I}$}},}\ }\href {\doibase
  10.1063/1.1708493} {\bibfield  {journal} {\bibinfo  {journal} {Journal of
  Applied Physics}\ }\textbf {\bibinfo {volume} {37}},\ \bibinfo {pages}
  {1404–1405} (\bibinfo {year} {1966})}\BibitemShut {NoStop}%
\bibitem [{\citenamefont {Schmid}\ and\ \citenamefont
  {Schwarzmüller}(1976)}]{Schmid_SchwarzmuLler_1976}%
  \BibitemOpen
  \bibfield  {author} {\bibinfo {author} {\bibfnamefont {H.}~\bibnamefont
  {Schmid}}\ and\ \bibinfo {author} {\bibfnamefont {J.}~\bibnamefont
  {Schwarzmüller}},\ }\bibfield  {title} {\enquote {\bibinfo {title} {{Review
  of Ferroelectric Materials Usable for Passive Electro-Optic Alphanumeric
  Display Devices}},}\ }\href {\doibase 10.1080/00150197608241996} {\bibfield
  {journal} {\bibinfo  {journal} {Ferroelectrics}\ }\textbf {\bibinfo {volume}
  {10}},\ \bibinfo {pages} {283–293} (\bibinfo {year} {1976})}\BibitemShut
  {NoStop}%
\bibitem [{\citenamefont {Schmid}\ \emph {et~al.}(1980)\citenamefont {Schmid},
  \citenamefont {Genequand}, \citenamefont {Pouilly},\ and\ \citenamefont
  {Chan}}]{Schmid_Genequand_Pouilly_Chan_1980}%
  \BibitemOpen
  \bibfield  {author} {\bibinfo {author} {\bibfnamefont {H.}~\bibnamefont
  {Schmid}}, \bibinfo {author} {\bibfnamefont {P.}~\bibnamefont {Genequand}},
  \bibinfo {author} {\bibfnamefont {G.}~\bibnamefont {Pouilly}}, \ and\
  \bibinfo {author} {\bibfnamefont {P.}~\bibnamefont {Chan}},\ }\bibfield
  {title} {\enquote {\bibinfo {title} {{Pyroelectricity of $\textrm{Fe-I}$ and
  \textrm{Cu-Cl} boracite}},}\ }\href@noop {} {\bibfield  {journal} {\bibinfo
  {journal} {Ferroelectrics}\ }\textbf {\bibinfo {volume} {25}},\ \bibinfo
  {pages} {539–542} (\bibinfo {year} {1980})}\BibitemShut {NoStop}%
\bibitem [{\citenamefont {Smith}, \citenamefont {Rosar},\ and\ \citenamefont
  {Shaulov}(1981)}]{Smith_Rosar_Shaulov_1981}%
  \BibitemOpen
  \bibfield  {author} {\bibinfo {author} {\bibfnamefont {W.~A.}\ \bibnamefont
  {Smith}}, \bibinfo {author} {\bibfnamefont {M.~E.}\ \bibnamefont {Rosar}}, \
  and\ \bibinfo {author} {\bibfnamefont {A.}~\bibnamefont {Shaulov}},\
  }\bibfield  {title} {\enquote {\bibinfo {title} {{Analysis of pyroelectric
  and dielectric measurements on boracites}},}\ }\href {\doibase
  10.1080/00150198108218155} {\bibfield  {journal} {\bibinfo  {journal}
  {Ferroelectrics}\ }\textbf {\bibinfo {volume} {36}},\ \bibinfo {pages}
  {467–470} (\bibinfo {year} {1981})}\BibitemShut {NoStop}%
\bibitem [{\citenamefont {Whatmore}, \citenamefont {Herbert},\ and\
  \citenamefont {Ainger}(1980)}]{Whatmore_Herbert_Ainger_1980}%
  \BibitemOpen
  \bibfield  {author} {\bibinfo {author} {\bibfnamefont {R.~W.}\ \bibnamefont
  {Whatmore}}, \bibinfo {author} {\bibfnamefont {J.~M.}\ \bibnamefont
  {Herbert}}, \ and\ \bibinfo {author} {\bibfnamefont {F.~W.}\ \bibnamefont
  {Ainger}},\ }\bibfield  {title} {\enquote {\bibinfo {title} {{Recent
  developments in ferroelectrics for infrared detectors}},}\ }\href {\doibase
  10.1002/pssa.2210610106} {\bibfield  {journal} {\bibinfo  {journal} {Physica
  Status Solidi (a)}\ }\textbf {\bibinfo {volume} {61}},\ \bibinfo {pages}
  {73–80} (\bibinfo {year} {1980})}\BibitemShut {NoStop}%
\bibitem [{\citenamefont {Schmid}(1965)}]{Schmid_1965}%
  \BibitemOpen
  \bibfield  {author} {\bibinfo {author} {\bibfnamefont {H.}~\bibnamefont
  {Schmid}},\ }\bibfield  {title} {\enquote {\bibinfo {title} {{Die synthese
  von boraziten mit hilfe chemischer transportreaktionen}},}\ }\href {\doibase
  10.1016/0022-3697(65)90185-X} {\bibfield  {journal} {\bibinfo  {journal}
  {Journal of Physics and Chemistry of Solids}\ }\textbf {\bibinfo {volume}
  {26}},\ \bibinfo {pages} {973–976} (\bibinfo {year} {1965})}\BibitemShut
  {NoStop}%
\bibitem [{\citenamefont {Whatmore}, \citenamefont {Brierley},\ and\
  \citenamefont {Ainger}(1980)}]{Whatmore_Brierley_Ainger_1980}%
  \BibitemOpen
  \bibfield  {author} {\bibinfo {author} {\bibfnamefont {R.~W.}\ \bibnamefont
  {Whatmore}}, \bibinfo {author} {\bibfnamefont {C.~J.}\ \bibnamefont
  {Brierley}}, \ and\ \bibinfo {author} {\bibfnamefont {F.~W.}\ \bibnamefont
  {Ainger}},\ }\bibfield  {title} {\enquote {\bibinfo {title} {{Nucleation
  control during the growth of boracite single crystals}},}\ }\href {\doibase
  10.1080/00150198008227101} {\bibfield  {journal} {\bibinfo  {journal}
  {Ferroelectrics}\ }\textbf {\bibinfo {volume} {28}},\ \bibinfo {pages}
  {329–332} (\bibinfo {year} {1980})}\BibitemShut {NoStop}%
\bibitem [{\citenamefont {Catalan}\ \emph {et~al.}(2012)\citenamefont
  {Catalan}, \citenamefont {Seidel}, \citenamefont {Ramesh},\ and\
  \citenamefont {Scott}}]{Catalan_Seidel_Ramesh_Scott_2012}%
  \BibitemOpen
  \bibfield  {author} {\bibinfo {author} {\bibfnamefont {G.}~\bibnamefont
  {Catalan}}, \bibinfo {author} {\bibfnamefont {J.}~\bibnamefont {Seidel}},
  \bibinfo {author} {\bibfnamefont {R.}~\bibnamefont {Ramesh}}, \ and\ \bibinfo
  {author} {\bibfnamefont {J.}~\bibnamefont {Scott}},\ }\bibfield  {title}
  {\enquote {\bibinfo {title} {{Domain wall nanoelectronics}},}\ }\href
  {\doibase 10.1103/RevModPhys.84.119} {\bibfield  {journal} {\bibinfo
  {journal} {Reviews of Modern Physics}\ }\textbf {\bibinfo {volume} {84}},\
  \bibinfo {pages} {119–156} (\bibinfo {year} {2012})}\BibitemShut {NoStop}%
\bibitem [{\citenamefont {Evans}\ \emph {et~al.}(2020)\citenamefont {Evans},
  \citenamefont {Cochard}, \citenamefont {McQuaid}, \citenamefont {Cano},
  \citenamefont {Gregg},\ and\ \citenamefont
  {Meier}}]{Evans_Cochard_McQuaid_Cano_Gregg_Meier_2020}%
  \BibitemOpen
  \bibfield  {author} {\bibinfo {author} {\bibfnamefont {D.~M.}\ \bibnamefont
  {Evans}}, \bibinfo {author} {\bibfnamefont {C.}~\bibnamefont {Cochard}},
  \bibinfo {author} {\bibfnamefont {R.~G.~P.}\ \bibnamefont {McQuaid}},
  \bibinfo {author} {\bibfnamefont {A.}~\bibnamefont {Cano}}, \bibinfo {author}
  {\bibfnamefont {J.~M.}\ \bibnamefont {Gregg}}, \ and\ \bibinfo {author}
  {\bibfnamefont {D.}~\bibnamefont {Meier}},\ }\enquote {\bibinfo {title}
  {{Improper Ferroelectric Domain Walls}},}\ in\ \href {\doibase
  10.1093/oso/9780198862499.003.0006} {\emph {\bibinfo {booktitle} {Domain
  Walls}}}\ (\bibinfo  {publisher} {Oxford University Press},\ \bibinfo {year}
  {2020})\ p.\ \bibinfo {pages} {129–151}\BibitemShut {NoStop}%
\bibitem [{\citenamefont {Feng}\ \emph {et~al.}(2018)\citenamefont {Feng},
  \citenamefont {Xu}, \citenamefont {Bellaiche},\ and\ \citenamefont
  {Xiang}}]{Feng_Xu_Bellaiche_Xiang_2018}%
  \BibitemOpen
  \bibfield  {author} {\bibinfo {author} {\bibfnamefont {J.~S.}\ \bibnamefont
  {Feng}}, \bibinfo {author} {\bibfnamefont {K.}~\bibnamefont {Xu}}, \bibinfo
  {author} {\bibfnamefont {L.}~\bibnamefont {Bellaiche}}, \ and\ \bibinfo
  {author} {\bibfnamefont {H.~J.}\ \bibnamefont {Xiang}},\ }\bibfield  {title}
  {\enquote {\bibinfo {title} {{Designing switchable near room-temperature
  multiferroics via the discovery of a novel magnetoelectric coupling}},}\
  }\href {\doibase 10.1088/1367-2630/aabed3} {\bibfield  {journal} {\bibinfo
  {journal} {New Journal of Physics}\ }\textbf {\bibinfo {volume} {20}},\
  \bibinfo {pages} {053025} (\bibinfo {year} {2018})}\BibitemShut {NoStop}%
\bibitem [{\citenamefont {Guy}\ \emph {et~al.}(2021)\citenamefont {Guy},
  \citenamefont {Cochard}, \citenamefont {Aguado-Puente}, \citenamefont
  {Soergel}, \citenamefont {Whatmore}, \citenamefont {Conroy}, \citenamefont
  {Moore}, \citenamefont {Courtney}, \citenamefont {Harvey}, \citenamefont
  {Bangert}, \citenamefont {Kumar}, \citenamefont {McQuaid},\ and\
  \citenamefont {Gregg}}]{Guy2021}%
  \BibitemOpen
  \bibfield  {author} {\bibinfo {author} {\bibfnamefont {J.~G.}\ \bibnamefont
  {Guy}}, \bibinfo {author} {\bibfnamefont {C.}~\bibnamefont {Cochard}},
  \bibinfo {author} {\bibfnamefont {P.}~\bibnamefont {Aguado-Puente}}, \bibinfo
  {author} {\bibfnamefont {E.}~\bibnamefont {Soergel}}, \bibinfo {author}
  {\bibfnamefont {R.~W.}\ \bibnamefont {Whatmore}}, \bibinfo {author}
  {\bibfnamefont {M.}~\bibnamefont {Conroy}}, \bibinfo {author} {\bibfnamefont
  {K.}~\bibnamefont {Moore}}, \bibinfo {author} {\bibfnamefont
  {E.}~\bibnamefont {Courtney}}, \bibinfo {author} {\bibfnamefont
  {A.}~\bibnamefont {Harvey}}, \bibinfo {author} {\bibfnamefont
  {U.}~\bibnamefont {Bangert}}, \bibinfo {author} {\bibfnamefont
  {A.}~\bibnamefont {Kumar}}, \bibinfo {author} {\bibfnamefont {R.~G.}\
  \bibnamefont {McQuaid}}, \ and\ \bibinfo {author} {\bibfnamefont {J.~M.}\
  \bibnamefont {Gregg}},\ }\bibfield  {title} {\enquote {\bibinfo {title}
  {{Anomalous Motion of Charged Domain Walls and Associated Negative
  Capacitance in Copper–Chlorine Boracite}},}\ }\href {\doibase
  10.1002/adma.202008068} {\bibfield  {journal} {\bibinfo  {journal} {Advanced
  Materials}\ }\textbf {\bibinfo {volume} {33}},\ \bibinfo {pages} {2008068}
  (\bibinfo {year} {2021})}\BibitemShut {NoStop}%
\bibitem [{\citenamefont {McQuaid}\ \emph {et~al.}(2017)\citenamefont
  {McQuaid}, \citenamefont {Campbell}, \citenamefont {Whatmore}, \citenamefont
  {Kumar},\ and\ \citenamefont
  {Gregg}}]{McQuaid_Campbell_Whatmore_Kumar_Gregg_2017}%
  \BibitemOpen
  \bibfield  {author} {\bibinfo {author} {\bibfnamefont {R.~G.}\ \bibnamefont
  {McQuaid}}, \bibinfo {author} {\bibfnamefont {M.~P.}\ \bibnamefont
  {Campbell}}, \bibinfo {author} {\bibfnamefont {R.~W.}\ \bibnamefont
  {Whatmore}}, \bibinfo {author} {\bibfnamefont {A.}~\bibnamefont {Kumar}}, \
  and\ \bibinfo {author} {\bibfnamefont {J.~M.}\ \bibnamefont {Gregg}},\
  }\bibfield  {title} {\enquote {\bibinfo {title} {{Injection and controlled
  motion of conducting domain walls in improper ferroelectric Cu-Cl
  boracite}},}\ }\href {\doibase 10.1038/ncomms15105} {\bibfield  {journal}
  {\bibinfo  {journal} {Nature Communications}\ }\textbf {\bibinfo {volume}
  {8}},\ \bibinfo {pages} {15105} (\bibinfo {year} {2017})}\BibitemShut
  {NoStop}%
\bibitem [{\citenamefont {Seidel}(2012)}]{Seidel_2012}%
  \BibitemOpen
  \bibfield  {author} {\bibinfo {author} {\bibfnamefont {J.}~\bibnamefont
  {Seidel}},\ }\bibfield  {title} {\enquote {\bibinfo {title} {{Domain walls as
  nanoscale functional elements}},}\ }\href {\doibase 10.1021/jz3011223}
  {\bibfield  {journal} {\bibinfo  {journal} {Journal of Physical Chemistry
  Letters}\ }\textbf {\bibinfo {volume} {3}},\ \bibinfo {pages} {2905–2909}
  (\bibinfo {year} {2012})}\BibitemShut {NoStop}%
\bibitem [{\citenamefont {Whyte}\ and\ \citenamefont
  {Gregg}(2015)}]{Whyte_Gregg_2015}%
  \BibitemOpen
  \bibfield  {author} {\bibinfo {author} {\bibfnamefont {J.~R.}\ \bibnamefont
  {Whyte}}\ and\ \bibinfo {author} {\bibfnamefont {J.~M.}\ \bibnamefont
  {Gregg}},\ }\bibfield  {title} {\enquote {\bibinfo {title} {{A diode for
  ferroelectric domain-wall motion}},}\ }\href {\doibase 10.1038/ncomms8361}
  {\bibfield  {journal} {\bibinfo  {journal} {Nature Communications}\ }\textbf
  {\bibinfo {volume} {6}},\ \bibinfo {pages} {1–5} (\bibinfo {year}
  {2015})}\BibitemShut {NoStop}%
\bibitem [{\citenamefont {Fernandez-Posada}\ \emph {et~al.}(2021)\citenamefont
  {Fernandez-Posada}, \citenamefont {Cochard}, \citenamefont {Gregg},
  \citenamefont {Whatmore},\ and\ \citenamefont
  {Carpenter}}]{Fernandez-Posada_Cochard_Gregg_Whatmore_Carpenter_2021}%
  \BibitemOpen
  \bibfield  {author} {\bibinfo {author} {\bibfnamefont {C.~M.}\ \bibnamefont
  {Fernandez-Posada}}, \bibinfo {author} {\bibfnamefont {C.}~\bibnamefont
  {Cochard}}, \bibinfo {author} {\bibfnamefont {J.~M.}\ \bibnamefont {Gregg}},
  \bibinfo {author} {\bibfnamefont {R.~W.}\ \bibnamefont {Whatmore}}, \ and\
  \bibinfo {author} {\bibfnamefont {M.~A.}\ \bibnamefont {Carpenter}},\
  }\bibfield  {title} {\enquote {\bibinfo {title} {{Order–disorder,
  ferroelasticity and mobility of domain walls in multiferroic Cu–Cl
  boracite}},}\ }\href {\doibase 10.1088/1361-648X/abcb0f} {\bibfield
  {journal} {\bibinfo  {journal} {Journal of Physics: Condensed Matter}\
  }\textbf {\bibinfo {volume} {33}},\ \bibinfo {pages} {095402} (\bibinfo
  {year} {2021})}\BibitemShut {NoStop}%
\bibitem [{\citenamefont {Granzow}(2017)}]{Granzow_2017}%
  \BibitemOpen
  \bibfield  {author} {\bibinfo {author} {\bibfnamefont {T.}~\bibnamefont
  {Granzow}},\ }\bibfield  {title} {\enquote {\bibinfo {title}
  {{Polaron-mediated low-frequency dielectric anomaly in reduced
  $\textrm{LiNbO}_3\textrm{:Ti}$}},}\ }\href {\doibase 10.1063/1.4990389}
  {\bibfield  {journal} {\bibinfo  {journal} {Applied Physics Letters}\
  }\textbf {\bibinfo {volume} {111}},\ \bibinfo {pages} {022903} (\bibinfo
  {year} {2017})}\BibitemShut {NoStop}%
\bibitem [{\citenamefont {Havriliak}\ and\ \citenamefont
  {Negami}(1967)}]{Havriliak_Negami_1967}%
  \BibitemOpen
  \bibfield  {author} {\bibinfo {author} {\bibfnamefont {S.}~\bibnamefont
  {Havriliak}}\ and\ \bibinfo {author} {\bibfnamefont {S.}~\bibnamefont
  {Negami}},\ }\bibfield  {title} {\enquote {\bibinfo {title} {{A complex plane
  representation of dielectric and mechanical relaxation processes in some
  polymers}},}\ }\href {\doibase 10.1016/0032-3861(67)90021-3} {\bibfield
  {journal} {\bibinfo  {journal} {Polymer}\ }\textbf {\bibinfo {volume} {8}},\
  \bibinfo {pages} {161–210} (\bibinfo {year} {1967})}\BibitemShut {NoStop}%
\bibitem [{\citenamefont {Cochard}\ \emph {et~al.}(2016)\citenamefont
  {Cochard}, \citenamefont {Bril}, \citenamefont {Guedes},\ and\ \citenamefont
  {Janolin}}]{Cochard_Bril_Guedes_Janolin_2016}%
  \BibitemOpen
  \bibfield  {author} {\bibinfo {author} {\bibfnamefont {C.}~\bibnamefont
  {Cochard}}, \bibinfo {author} {\bibfnamefont {X.}~\bibnamefont {Bril}},
  \bibinfo {author} {\bibfnamefont {O.}~\bibnamefont {Guedes}}, \ and\ \bibinfo
  {author} {\bibfnamefont {P.-E.}\ \bibnamefont {Janolin}},\ }\bibfield
  {title} {\enquote {\bibinfo {title} {{Interpretation of Polar Orders Based on
  Electric Characterizations: Example of Pb(Yb$_{1/2}$Nb$_{1/2}$)O$_3$
  -PbTiO$_3$ Solid Solution}},}\ }\href {\doibase 10.1007/s11664-016-4758-0}
  {\bibfield  {journal} {\bibinfo  {journal} {Journal of Electronic Materials}\
  }\textbf {\bibinfo {volume} {45}},\ \bibinfo {pages} {6005–6011} (\bibinfo
  {year} {2016})}\BibitemShut {NoStop}%
\bibitem [{\citenamefont {Elissalde}\ and\ \citenamefont
  {Ravez}(2001)}]{Elissalde_Ravez_2001}%
  \BibitemOpen
  \bibfield  {author} {\bibinfo {author} {\bibfnamefont {C.}~\bibnamefont
  {Elissalde}}\ and\ \bibinfo {author} {\bibfnamefont {J.}~\bibnamefont
  {Ravez}},\ }\bibfield  {title} {\enquote {\bibinfo {title} {{Ferroelectric
  ceramics: Defects and dielectric relaxations}},}\ }\href {\doibase
  10.1039/b010117f} {\bibfield  {journal} {\bibinfo  {journal} {Journal of
  Materials Chemistry}\ }\textbf {\bibinfo {volume} {11}},\ \bibinfo {pages}
  {1957–1967} (\bibinfo {year} {2001})}\BibitemShut {NoStop}%
\bibitem [{\citenamefont {Nataf}\ \emph {et~al.}(2016)\citenamefont {Nataf},
  \citenamefont {Aktas}, \citenamefont {Granzow},\ and\ \citenamefont
  {Salje}}]{Nataf_Aktas_Granzow_Salje_2016}%
  \BibitemOpen
  \bibfield  {author} {\bibinfo {author} {\bibfnamefont {G.~F.}\ \bibnamefont
  {Nataf}}, \bibinfo {author} {\bibfnamefont {O.}~\bibnamefont {Aktas}},
  \bibinfo {author} {\bibfnamefont {T.}~\bibnamefont {Granzow}}, \ and\
  \bibinfo {author} {\bibfnamefont {E.~K.}\ \bibnamefont {Salje}},\ }\bibfield
  {title} {\enquote {\bibinfo {title} {{Influence of defects and domain walls
  on dielectric and mechanical resonances in LiNbO$_3$}},}\ }\href {\doibase
  10.1088/0953-8984/28/1/015901} {\bibfield  {journal} {\bibinfo  {journal}
  {Journal of Physics: Condensed Matter}\ }\textbf {\bibinfo {volume} {28}},\
  \bibinfo {pages} {015901} (\bibinfo {year} {2016})}\BibitemShut {NoStop}%
\bibitem [{\citenamefont {Pramanick}\ \emph {et~al.}(2012)\citenamefont
  {Pramanick}, \citenamefont {Prewitt}, \citenamefont {Forrester},\ and\
  \citenamefont {Jones}}]{Pramanick_Prewitt_Forrester_Jones_2012}%
  \BibitemOpen
  \bibfield  {author} {\bibinfo {author} {\bibfnamefont {A.}~\bibnamefont
  {Pramanick}}, \bibinfo {author} {\bibfnamefont {A.~D.}\ \bibnamefont
  {Prewitt}}, \bibinfo {author} {\bibfnamefont {J.~S.}\ \bibnamefont
  {Forrester}}, \ and\ \bibinfo {author} {\bibfnamefont {J.~L.}\ \bibnamefont
  {Jones}},\ }\bibfield  {title} {\enquote {\bibinfo {title} {{Domains, domain
  walls and defects in perovskite ferroelectric oxides: A review of present
  understanding and recent contributions}},}\ }\href {\doibase
  10.1080/10408436.2012.686891} {\bibfield  {journal} {\bibinfo  {journal}
  {Critical Reviews in Solid State and Materials Sciences}\ }\textbf {\bibinfo
  {volume} {37}},\ \bibinfo {pages} {243–275} (\bibinfo {year}
  {2012})}\BibitemShut {NoStop}%
\bibitem [{\citenamefont {Akkopru-Akgun}(2019)}]{Akkopru-Akgun_2019}%
  \BibitemOpen
  \bibfield  {author} {\bibinfo {author} {\bibfnamefont {B.}~\bibnamefont
  {Akkopru-Akgun}},\ }\emph {\bibinfo {title} {{The Role of Defect Chemistry in
  DC Resistance Degradation of Lead Zirconate Titanate Thin Films}}},\
  \href@noop {} {Ph.D. thesis} (\bibinfo {year} {2019})\BibitemShut {NoStop}%
\bibitem [{\citenamefont {Kim}\ \emph {et~al.}(1981)\citenamefont {Kim},
  \citenamefont {Somoano}, \citenamefont {{Lowe Ma}}, \citenamefont {Coleman},\
  and\ \citenamefont {Moopen}}]{Kim_Somoano_LoweMa_Coleman_Moopen_1981}%
  \BibitemOpen
  \bibfield  {author} {\bibinfo {author} {\bibfnamefont {Q.}~\bibnamefont
  {Kim}}, \bibinfo {author} {\bibfnamefont {R.}~\bibnamefont {Somoano}},
  \bibinfo {author} {\bibfnamefont {C.}~\bibnamefont {{Lowe Ma}}}, \bibinfo
  {author} {\bibfnamefont {L.~B.}\ \bibnamefont {Coleman}}, \ and\ \bibinfo
  {author} {\bibfnamefont {A.}~\bibnamefont {Moopen}},\ }\bibfield  {title}
  {\enquote {\bibinfo {title} {{Studies of defects in improper
  ferroelectrics}},}\ }\href {\doibase 10.1080/00150198108218147} {\bibfield
  {journal} {\bibinfo  {journal} {Ferroelectrics}\ }\textbf {\bibinfo {volume}
  {36}},\ \bibinfo {pages} {435--438} (\bibinfo {year} {1981})}\BibitemShut
  {NoStop}%
\bibitem [{\citenamefont {Gridnev}(2002)}]{Gridnev2002}%
  \BibitemOpen
  \bibfield  {author} {\bibinfo {author} {\bibfnamefont {S.~A.}\ \bibnamefont
  {Gridnev}},\ }\bibfield  {title} {\enquote {\bibinfo {title} {{Dielectric
  relaxation in disordered polar dielectrics}},}\ }\href {\doibase
  10.1080/00150190211452} {\bibfield  {journal} {\bibinfo  {journal}
  {Ferroelectrics}\ }\textbf {\bibinfo {volume} {266}},\ \bibinfo {pages}
  {171--209} (\bibinfo {year} {2002})}\BibitemShut {NoStop}%
\bibitem [{\citenamefont {El-Falaky}, \citenamefont {Guirguis},\ and\
  \citenamefont {Abd El-Aal}(2012)}]{El-Falaky_Guirguis_AbdEl-Aal_2012}%
  \BibitemOpen
  \bibfield  {author} {\bibinfo {author} {\bibfnamefont {G.~E.}\ \bibnamefont
  {El-Falaky}}, \bibinfo {author} {\bibfnamefont {O.~W.}\ \bibnamefont
  {Guirguis}}, \ and\ \bibinfo {author} {\bibfnamefont {N.~S.}\ \bibnamefont
  {Abd El-Aal}},\ }\bibfield  {title} {\enquote {\bibinfo {title} {{A.C.
  conductivity and relaxation dynamics in zinc–borate glasses}},}\ }\href
  {\doibase 10.1016/j.pnsc.2012.03.012} {\bibfield  {journal} {\bibinfo
  {journal} {Progress in Natural Science: Materials International}\ }\textbf
  {\bibinfo {volume} {22}},\ \bibinfo {pages} {86–93} (\bibinfo {year}
  {2012})}\BibitemShut {NoStop}%
\bibitem [{\citenamefont {Sapper}\ \emph {et~al.}(2014)\citenamefont {Sapper},
  \citenamefont {Dittmer}, \citenamefont {Damjanovic}, \citenamefont {Erdem},
  \citenamefont {Keeble}, \citenamefont {Jo}, \citenamefont {Granzow},\ and\
  \citenamefont
  {Rödel}}]{Sapper_Dittmer_Damjanovic_Erdem_Keeble_Jo_Granzow_Rodel_2014}%
  \BibitemOpen
  \bibfield  {author} {\bibinfo {author} {\bibfnamefont {E.}~\bibnamefont
  {Sapper}}, \bibinfo {author} {\bibfnamefont {R.}~\bibnamefont {Dittmer}},
  \bibinfo {author} {\bibfnamefont {D.}~\bibnamefont {Damjanovic}}, \bibinfo
  {author} {\bibfnamefont {E.}~\bibnamefont {Erdem}}, \bibinfo {author}
  {\bibfnamefont {D.~J.}\ \bibnamefont {Keeble}}, \bibinfo {author}
  {\bibfnamefont {W.}~\bibnamefont {Jo}}, \bibinfo {author} {\bibfnamefont
  {T.}~\bibnamefont {Granzow}}, \ and\ \bibinfo {author} {\bibfnamefont
  {J.}~\bibnamefont {Rödel}},\ }\bibfield  {title} {\enquote {\bibinfo {title}
  {{Aging in the relaxor and ferroelectric state of Fe-doped
  (1-x)(Bi$_{1/2}$Na$_{1/2}$)TiO$_3$-xBaTiO$_3$ piezoelectric ceramics}},}\
  }\href {\doibase 10.1063/1.4894630} {\bibfield  {journal} {\bibinfo
  {journal} {Journal of Applied Physics}\ }\textbf {\bibinfo {volume} {116}},\
  \bibinfo {pages} {104102} (\bibinfo {year} {2014})}\BibitemShut {NoStop}%
\bibitem [{\citenamefont {Zhang}\ \emph {et~al.}(2008)\citenamefont {Zhang},
  \citenamefont {Erdem}, \citenamefont {Ren},\ and\ \citenamefont
  {Eichel}}]{Zhang_Erdem_Ren_Eichel_2008}%
  \BibitemOpen
  \bibfield  {author} {\bibinfo {author} {\bibfnamefont {L.}~\bibnamefont
  {Zhang}}, \bibinfo {author} {\bibfnamefont {E.}~\bibnamefont {Erdem}},
  \bibinfo {author} {\bibfnamefont {X.}~\bibnamefont {Ren}}, \ and\ \bibinfo
  {author} {\bibfnamefont {R.-A.}\ \bibnamefont {Eichel}},\ }\bibfield  {title}
  {\enquote {\bibinfo {title} {{Reorientation of (Mn$_{Ti}''−$\VO) defect
  dipoles in acceptor-modified BaTiO$_3$ single crystals: An electron
  paramagnetic resonance study}},}\ }\href {\doibase 10.1063/1.3006327}
  {\bibfield  {journal} {\bibinfo  {journal} {Applied Physics Letters}\
  }\textbf {\bibinfo {volume} {93}},\ \bibinfo {pages} {202901} (\bibinfo
  {year} {2008})}\BibitemShut {NoStop}%
\bibitem [{\citenamefont {Eichel}\ \emph {et~al.}(2008)\citenamefont {Eichel},
  \citenamefont {Erhart}, \citenamefont {Träskelin}, \citenamefont {Albe},
  \citenamefont {Kungl},\ and\ \citenamefont
  {Hoffmann}}]{Eichel_Erhart_Traskelin_Albe_Kungl_Hoffmann_2008}%
  \BibitemOpen
  \bibfield  {author} {\bibinfo {author} {\bibfnamefont {R.~A.}\ \bibnamefont
  {Eichel}}, \bibinfo {author} {\bibfnamefont {P.}~\bibnamefont {Erhart}},
  \bibinfo {author} {\bibfnamefont {P.}~\bibnamefont {Träskelin}}, \bibinfo
  {author} {\bibfnamefont {K.}~\bibnamefont {Albe}}, \bibinfo {author}
  {\bibfnamefont {H.}~\bibnamefont {Kungl}}, \ and\ \bibinfo {author}
  {\bibfnamefont {M.~J.}\ \bibnamefont {Hoffmann}},\ }\bibfield  {title}
  {\enquote {\bibinfo {title} {{Defect-dipole formation in copper-doped PbTiO3
  ferroelectrics}},}\ }\href {\doibase 10.1103/PhysRevLett.100.095504}
  {\bibfield  {journal} {\bibinfo  {journal} {Physical Review Letters}\
  }\textbf {\bibinfo {volume} {100}},\ \bibinfo {pages} {1–4} (\bibinfo
  {year} {2008})}\BibitemShut {NoStop}%
\bibitem [{\citenamefont {Erhart}\ \emph {et~al.}(2007)\citenamefont {Erhart},
  \citenamefont {Eichel}, \citenamefont {Träskelin},\ and\ \citenamefont
  {Albe}}]{Erhart_Eichel_Traskelin_Albe_2007}%
  \BibitemOpen
  \bibfield  {author} {\bibinfo {author} {\bibfnamefont {P.}~\bibnamefont
  {Erhart}}, \bibinfo {author} {\bibfnamefont {R.~A.}\ \bibnamefont {Eichel}},
  \bibinfo {author} {\bibfnamefont {P.}~\bibnamefont {Träskelin}}, \ and\
  \bibinfo {author} {\bibfnamefont {K.}~\bibnamefont {Albe}},\ }\bibfield
  {title} {\enquote {\bibinfo {title} {{Association of oxygen vacancies with
  impurity metal ions in lead titanate}},}\ }\href {\doibase
  10.1103/PhysRevB.76.174116} {\bibfield  {journal} {\bibinfo  {journal}
  {Physical Review B - Condensed Matter and Materials Physics}\ }\textbf
  {\bibinfo {volume} {76}},\ \bibinfo {pages} {1–12} (\bibinfo {year}
  {2007})}\BibitemShut {NoStop}%
\bibitem [{\citenamefont {Schmid}\ and\ \citenamefont
  {Petermann}(1977)}]{Schmid_Petermann_1977}%
  \BibitemOpen
  \bibfield  {author} {\bibinfo {author} {\bibfnamefont {H.}~\bibnamefont
  {Schmid}}\ and\ \bibinfo {author} {\bibfnamefont {L.~A.}\ \bibnamefont
  {Petermann}},\ }\bibfield  {title} {\enquote {\bibinfo {title} {{Dielectric
  constant and electric resistivity of copper chlorine boracite,
  Cu$_3$B$_7$O$_{13}$Cl (Cu-Cl-B)}},}\ }\href@noop {} {\bibfield  {journal}
  {\bibinfo  {journal} {phys. stat. sol. (a)}\ }\textbf {\bibinfo {volume}
  {41}},\ \bibinfo {pages} {K147} (\bibinfo {year} {1977})}\BibitemShut
  {NoStop}%
\bibitem [{\citenamefont {Connor}\ \emph {et~al.}(2021)\citenamefont {Connor},
  \citenamefont {Smaha}, \citenamefont {Li}, \citenamefont {Gold-Parker},
  \citenamefont {Heyer}, \citenamefont {Toney}, \citenamefont {Lee},\ and\
  \citenamefont
  {Karunadasa}}]{Connor_Smaha_Li_Gold-Parker_Heyer_Toney_Lee_Karunadasa_2021}%
  \BibitemOpen
  \bibfield  {author} {\bibinfo {author} {\bibfnamefont {B.~A.}\ \bibnamefont
  {Connor}}, \bibinfo {author} {\bibfnamefont {R.~W.}\ \bibnamefont {Smaha}},
  \bibinfo {author} {\bibfnamefont {J.}~\bibnamefont {Li}}, \bibinfo {author}
  {\bibfnamefont {A.}~\bibnamefont {Gold-Parker}}, \bibinfo {author}
  {\bibfnamefont {A.~J.}\ \bibnamefont {Heyer}}, \bibinfo {author}
  {\bibfnamefont {M.~F.}\ \bibnamefont {Toney}}, \bibinfo {author}
  {\bibfnamefont {Y.~S.}\ \bibnamefont {Lee}}, \ and\ \bibinfo {author}
  {\bibfnamefont {H.~I.}\ \bibnamefont {Karunadasa}},\ }\bibfield  {title}
  {\enquote {\bibinfo {title} {{ Alloying a single and a double perovskite: a
  Cu +/2+ mixed-valence layered halide perovskite with strong optical
  absorption }},}\ }\href {\doibase 10.1039/d1sc01159f} {\bibfield  {journal}
  {\bibinfo  {journal} {Chemical Science}\ }\textbf {\bibinfo {volume} {12}},\
  \bibinfo {pages} {8689–8697} (\bibinfo {year} {2021})}\BibitemShut
  {NoStop}%
\bibitem [{\citenamefont {Eliseev}\ \emph {et~al.}(2011)\citenamefont
  {Eliseev}, \citenamefont {Morozovska}, \citenamefont {Svechnikov},
  \citenamefont {Gopalan},\ and\ \citenamefont
  {Shur}}]{Eliseev_Morozovska_Svechnikov_Gopalan_Shur_2011}%
  \BibitemOpen
  \bibfield  {author} {\bibinfo {author} {\bibfnamefont {E.~A.}\ \bibnamefont
  {Eliseev}}, \bibinfo {author} {\bibfnamefont {A.~N.}\ \bibnamefont
  {Morozovska}}, \bibinfo {author} {\bibfnamefont {G.~S.}\ \bibnamefont
  {Svechnikov}}, \bibinfo {author} {\bibfnamefont {V.}~\bibnamefont {Gopalan}},
  \ and\ \bibinfo {author} {\bibfnamefont {V.~Y.}\ \bibnamefont {Shur}},\
  }\bibfield  {title} {\enquote {\bibinfo {title} {{Static conductivity of
  charged domain walls in uniaxial ferroelectric semiconductors}},}\ }\href
  {\doibase 10.1103/PhysRevB.83.235313} {\bibfield  {journal} {\bibinfo
  {journal} {Physical Review B}\ }\textbf {\bibinfo {volume} {83}},\ \bibinfo
  {pages} {1–8} (\bibinfo {year} {2011})}\BibitemShut {NoStop}%
\bibitem [{\citenamefont {Liu}\ \emph {et~al.}(2015)\citenamefont {Liu},
  \citenamefont {Zheng}, \citenamefont {Koocher}, \citenamefont {Takenaka},
  \citenamefont {Wang},\ and\ \citenamefont
  {Rappe}}]{Liu_Zheng_Koocher_Takenaka_Wang_Rappe_2015}%
  \BibitemOpen
  \bibfield  {author} {\bibinfo {author} {\bibfnamefont {S.}~\bibnamefont
  {Liu}}, \bibinfo {author} {\bibfnamefont {F.}~\bibnamefont {Zheng}}, \bibinfo
  {author} {\bibfnamefont {N.~Z.}\ \bibnamefont {Koocher}}, \bibinfo {author}
  {\bibfnamefont {H.}~\bibnamefont {Takenaka}}, \bibinfo {author}
  {\bibfnamefont {F.}~\bibnamefont {Wang}}, \ and\ \bibinfo {author}
  {\bibfnamefont {A.~M.}\ \bibnamefont {Rappe}},\ }\bibfield  {title} {\enquote
  {\bibinfo {title} {{Ferroelectric domain wall induced band gap reduction and
  charge separation in organometal halide perovskites}},}\ }\href {\doibase
  10.1021/jz502666j} {\bibfield  {journal} {\bibinfo  {journal} {Journal of
  Physical Chemistry Letters}\ }\textbf {\bibinfo {volume} {6}},\ \bibinfo
  {pages} {693–699} (\bibinfo {year} {2015})}\BibitemShut {NoStop}%
\bibitem [{\citenamefont {Nataf}\ \emph {et~al.}(2020)\citenamefont {Nataf},
  \citenamefont {Guennou}, \citenamefont {Gregg}, \citenamefont {Meier},
  \citenamefont {Hlinka}, \citenamefont {Salje},\ and\ \citenamefont
  {Kreisel}}]{Nataf_Guennou_Gregg_Meier_Hlinka_Salje_Kreisel_2020}%
  \BibitemOpen
  \bibfield  {author} {\bibinfo {author} {\bibfnamefont {G.~F.}\ \bibnamefont
  {Nataf}}, \bibinfo {author} {\bibfnamefont {M.}~\bibnamefont {Guennou}},
  \bibinfo {author} {\bibfnamefont {J.~M.}\ \bibnamefont {Gregg}}, \bibinfo
  {author} {\bibfnamefont {D.}~\bibnamefont {Meier}}, \bibinfo {author}
  {\bibfnamefont {J.}~\bibnamefont {Hlinka}}, \bibinfo {author} {\bibfnamefont
  {E.~K.}\ \bibnamefont {Salje}}, \ and\ \bibinfo {author} {\bibfnamefont
  {J.}~\bibnamefont {Kreisel}},\ }\bibfield  {title} {\enquote {\bibinfo
  {title} {{Domain-wall engineering and topological defects in ferroelectric
  and ferroelastic materials}},}\ }\href {\doibase 10.1038/s42254-020-0235-z}
  {\bibfield  {journal} {\bibinfo  {journal} {Nature Reviews Physics}\ }\textbf
  {\bibinfo {volume} {2}},\ \bibinfo {pages} {634–648} (\bibinfo {year}
  {2020})}\BibitemShut {NoStop}%
\end{thebibliography}%

\end{document}